\documentclass{article}

\PassOptionsToPackage{numbers, compress}{natbib}
\usepackage{natbib}
\usepackage[final]{nips_2018}

\usepackage{graphicx}
\usepackage[utf8]{inputenc} % allow utf-8 input
\usepackage[T1]{fontenc}    % use 8-bit T1 fonts
\usepackage{hyperref}       % hyperlinks
\usepackage{url}            % simple URL typesetting
\usepackage{booktabs}       % professional-quality tables
\usepackage{amsfonts}       % blackboard math symbols
\usepackage{nicefrac}       % compact symbols for 1/2, etc.
\usepackage{microtype}      % microtypography
\usepackage{caption}
\usepackage{wrapfig}
\usepackage{lipsum}

\title{Beyond the Leaderboard: Insight and Deployment Challenges to Address Research Problems}

\author{
  Adriënne M. Mendrik \\
  Netherlands eScience Center \\
  Science Park 140 (Matrix I) \\
  1098 XG Amsterdam, The Netherlands \\
  \texttt{a.mendrik@esciencecenter.nl} \\
  \And
  Stephen R.~Aylward \\
  Kitware, Inc. \\
  101 East Weaver Street, Suite G4 \\
  Carrboro, North Carolina, 27510 USA \\
}

\begin{document}
\maketitle
%\begin{abstract}...\end{abstract}
\section{Introduction}
In the medical image analysis field, organizing challenges with associated workshops at international conferences began in 2007 and has grown to include over 150 challenges \cite{Ginneken18}. Several of these challenges have had a major impact in the field, such as the Camelyon Challenge, which showed that deep learning algorithms can outperform a panel of 11 pathologists \cite{bejnordi17}. However, whereas well-designed challenges have the potential to unite and focus the field on creating solutions to important problems, poorly designed and documented challenges can equally impede a field and lead to pursuing incremental improvements in metric scores with no theoretic or clinical significance. 

The potential detriment of challenges inspired a critical assessment of challenges at the international Medical Image Computer and Computer-Assisted Intervention (MICCAI) conference and initiated a definition of best practices for challenge design \cite{Reinke18}. The main criticism raised is based on the observation that small changes to the underlying challenge data can drastically change the ranking order in the leaderboard. Related to this is the practice of leaderboard climbing \cite{Hardt15,whitehill17}, which is characterized by participants focusing on incrementally improving metric results rather than advancing science or solving the challenge's driving problem. In terms of challenge design, ways to avoid these criticisms are, for example, limiting the number of resubmissions, using a hold-out test set for the final leaderboard \cite{Dwork15}, or only updating the leaderboard in case of a significant change \cite{Blum15}. 

In this abstract we look beyond the leaderboard of a challenge and instead look at the conclusions that can be drawn from a challenge with respect to the research problem that it is addressing. Research study design is well described in other research areas, such as educational research \cite{Creswell02} and can be translated to challenge design when viewing challenges as research studies on algorithm performance that address a research problem. Based on the two main types of scientific research study design, we propose two main challenge types, which we think would benefit other research areas as well: 1) an "insight" challenge that is based on a \textit{qualitative} study design and 2) a "deployment" challenge that is based on a \textit{quantitative} study design. 

\section{Insight and Deployment Challenges}
Although a challenge in itself is a quantitative system of comparison, the design of its elements (i.e. data, truth, and metrics) can follow a qualitative study design. Like often in a research setting, circumstances can be sub-optimal in terms of generalization. In medical image analysis, bronze standards for truth \cite{Jannin02} are often the best that can be provided due to cost and time constraints as well as uncertainties regarding medical diagnostic categories. Therefore many challenges use small non-randomly selected datasets or in some cases incorporate indirect metrics. These uncertainties in truth, indirect metrics, and small non-randomly selected datasets are the hallmarks of a qualitative experiment. As qualitative experiments, the leaderboards of these challenges or subsequent tests cannot be used to draw quantitative conclusions about the research problem targeted by the challenge. Attempting to compute quantitative statistics from qualitative experiments is analogous to applying a t-test when the source data was not randomly sampled or does not have a normal distribution. 

Challenges that follow a qualitative design, however, can provide insights into a research problem \cite{Onwuegbuzie07, Creswell02}. For example, a cluster analysis of a qualitative challenge’s leaderboard (e.g., top 25\% of the algorithms) combined with knowledge about those algorithms’ methodologies can be used to identify which class of algorithms have a high potential to perform well for a class of problems. Additionally, qualitative challenges can be used to gain insight into how to achieve representative data, truth, and metrics for future challenges, including quantitative challenges. In qualitative research design this is called grounded theory \cite{Glaser68}. It is based on systematically analyzing qualitative data to discover hypotheses and population variables, and provide evidence that is grounded in the data. When design choices for insight challenges are openly described with respect to data sampling (e.g. purposive, convenience, etc), truth protocol and used metrics, and clearly follow from a research problem, purpose statement and research question, they can be used to draw conclusions that lead to new insights for research problems. These insights can in turn inspire new challenges addressing the same research problem from a different angle, clearly advancing in the direction of solving the problem or a certain part of the research problem. This is illustrated in Figure 1.

\begin{wrapfigure}{R}{0.45\textwidth}
%\begin{figure}[h!]
\begin{minipage}{.45\textwidth}
\centering
  \includegraphics[width=1.0\textwidth]{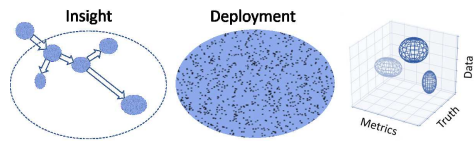}
  \caption{The dotted ellipse illustrates the problem space (statistical population) of a research problem, and how insight challenges cover a small part of this space, but can spark other insight challenges. Deployment challenges should provide data, truth, and metrics that are representative for the entire problem space to ensure generalization. The right figure illustrates how insight challenges can cover different parts of the high dimensional problem space associated with variations in data, truth, and metrics.}
\end{minipage}
%\end{figure}
\end{wrapfigure}

For deployment challenges, the goal is to identify an algorithm that successfully solves a research problem.  The data (randomly sampled from the statistical population and sufficiently large), truth, and metrics should be representative of the research problem. Additionally, beyond the leaderboard, a hold-out test should ultimately be used to test the hypothesis that an algorithm successfully solves the research problem, such that it could be deployed. 

Another thing to consider when looking beyond the leaderboard is the way algorithms are compared with each other. One of the most common practices is to test for statistically significant differences between methods; however, statistical significance does not capture generalizability or practical significance. The widespread inappropriate reliance on statistical significance has prompted many statistical journals to discourage its use \cite{Woolston15,Nuzzo14}. Generalizability (e.g., ANOVA methods \cite{Jagger08}) captures the potential sources of the differences between the two methods. Practical significance (e.g., scoring-guide scales \cite{gall01}) indicates if findings have any decision-making utility. In medicine, practical significance is often referred to as “clinical significance” to reflect the fact that small yet statistically significant differences may have no impact on treatment or outcome. For example, a small yet statistically significant improvement in tumor boundary estimation may not enable a reduction in the radiation therapy tumor margins \cite{Hojat04}.

Some challenges focus on increasing sample size in order to increase statistical significance; however, when looking beyond the leaderboard, such efforts are often marginalized due to data saturation. Data saturation is defined as encompassing insufficiencies in the data, metrics, or truth that diminish the statistical and practical significance as well as the generalizability that can be achieved as sample size is increased.  Admittedly, the concept of data saturation has been hotly debated \cite{Saunders18}. However, it is particularly relevant to current practices of (a) data augmentation, which may increase the variance in the data in a manner that disrupts the matching of population statistics; and (b) mechanical turk, which may provide more data at the cost of decreased quality of truth which, in turn, negates any benefit from increased sample size.

\section{Conclusion}
Although challenge leaderboards provide a quick overview of algorithm performance and can boost competition, we urge researchers to look beyond the leaderboard for challenge assessment. In this abstract, we make a distinction between insight and deployment challenges to address a research problem, and describe some additional considerations with respect to challenge design. 

%Please prepare submission files with paper size ``US Letter,'' and
%not, for example, ``A4.''

\medskip
\small
\end{document}